\documentclass[11pt]{article}
\usepackage{blois,epsfig}

\def\be{\begin{equation}}
\def\ee{\end{equation}}
\def\bea{\begin{eqnarray}}
\def\eea{\end{eqnarray}}

\def\spose#1{\hbox to 0pt{#1\hss}}
\def\simlt{\mathrel{\spose{\lower 3pt\hbox{$\mathchar"218$}}
     \raise 2.0pt\hbox{$\mathchar"13C$}}}
\def\simgt{\mathrel{\spose{\lower 3pt\hbox{$\mathchar"218$}}
     \raise 2.0pt\hbox{$\mathchar"13E$}}}
\def\lsim{\rlap{$<$}{\lower 1.0ex\hbox{$\sim$}}}
\def\gsim{\rlap{$>$}{\lower 1.0ex\hbox{$\sim$}}}

\def\kms{km~s$^{-1}$}

\def\eps@scaling{.95}
\def\epsscale#1{\gdef\eps@scaling{#1}}

\def\plotfiddle#1#2#3#4#5#6#7{\centering \leavevmode
    \vbox to#2{\rule{0pt}{#2}}
    \includegraphics{#1}}

\begin{document}
\vspace*{4cm}
\title{STIS Spectroscopy of PKS~0405-123}

\author{G. Williger$^1$, S. Heap$^2$, R. Weymann$^3$, R. Dav\'e$^4$, T. Tripp$^{5}$}

\address{$^1$ Dept of Physics \& Astronomy, Johns Hopkins U, Baltimore MD, USA\\
$^2$ Code 681, NASA Goddard Space Flight Center, Greenbelt MD, USA\\
$^3$ Obs Carnegie Inst of Washington, Pasadena CA, USA\\
$^4$ Steward Obs, U Arizona, Tucson AZ, USA\\
$^5$ Princeton U Obs, Princeton NJ; currently: Dept of Astronomy, U Massachusetts, Amherst MA, USA}

\maketitle\abstracts{
  We present results 
  for QSO PKS~0405--123 ($z=0.574$, $V=14.9$), as
  part of a STIS Investigation Definition Team (IDT) key project to
  study weak Ly$\alpha$ forest systems
  at low $z$.  We detect 59 (47) Ly$\alpha$ absorbers at
  4.0$\sigma$ significance to an 80\% completeness limit of column density $\log
  ({\rm N_{HI}})=13.3$ (13.1) for Doppler parameter $V_{Dop}=40$~\kms\ 
  over $0.002<z<0.423$ ($0.020<z<0.234$).
  We find 4 intervening O{\sc VI} systems, useful for
  studies of hot intergalactic gas.  We do not distinguish between
  metal and Ly$\alpha$-only systems in the following analysis.
The redshift density is consistent with previous measurements for
$\log {\rm N_{HI}}\geq 14.0$, but exhibits twice as many systems at
$13.1<\log {\rm N_{HI}}< 14.0$ compared to the
mean number density of
lines at $z<0.07$ toward 15 extragalactic objects. 
The difference possibly arises from cosmic variance.
The Doppler parameter distribution has $\langle V_{Dop} \rangle = 48
\pm 21$ km~s$^{-1}$; line blending possibly inflates the value.  We
find evidence for Ly$\alpha$-Ly$\alpha$ clustering in our sample on a
scale of $\Delta v \leq 250$ \kms , 
and there is evidence for a void at $0.032<z<0.081$ 
with probability of occurrance  $P=0.0005$.  
We find line-of-sight
velocity correlations of up to 250 \kms\ between Ly$\alpha$ absorbers
with $\log({\rm N_{HI}})\geq 13.1$ and 45 galaxies taken from the
literature and unpublished data at $0<z<0.47$; the transverse
distances cover up to 1.5 $h^{-1}_{70}$ Mpc in the local frame.  The
Ly$\alpha$-galaxy clustering is stronger for higher $\log({\rm
  N_{HI}})$ systems.
}

\section{Introduction}

PKS 0405--123 ($V=14.9$, $z=0.574$) is one of the brightest QSOs in
the sky, and illuminates a much longer span of the Ly$\alpha$ forest
than most of the other QSOs of similar or brighter magnitude.  It is
therefore a prime target for studies of the low $z$ Ly$\alpha$ forest
and metal absorbers
and
mid-latitude ($\ell=295^{\rm \, o} $, $b=-42^{\rm \, o}$) studies of Galactic absorption.  
We use $\sim 7$~\kms\ resolution STIS data to calculate the Ly$\alpha$
forest redshift density, Doppler parameter, clustering and void
statistics, and to cross-correlate the Ly$\alpha$ forest and field
galaxy redshifts.

\section{Observations, reductions and absorber sample}

PKS 0405--123 was observed with HST+STIS using grating E140M for ten
orbits (27208 sec) 
on 1999 Jan 24 and 1999 Mar 7. We used 
the $0.2''\times 0.06''$ slit for maximal
spectral purity.
The
data were reduced and extracted using STIS IDT team software at
Goddard Space Flight Center.  We constructed a continuum using a
combination of automated {\sc autovp} by R. Dav\'e and interactive
{\sc line\_norm} by D. Lindler, as the regions around emission lines
were problematic.  Absorption features were selected at the $4.0
\sigma$ significance level with a Gaussian filter, with half-widths of
10, 15, 20 pixels ($\sim 3$ \kms / pixel), and confirmed with a simple
equivalent width significance criterion for contiguous absorption
pixels.

Williger and R. Carswell (IOA, Cambridge) independently profile-fitted
the data with {\sc vpfit} (Webb 1987) using Voigt profiles convolved
with the STIS line spread
function, with largely consistent results.  We then
analyzed 1440 simulated Ly$\alpha$ lines to determine the 80\%
completeness limit, parametrized by $\log {\rm N_{HI}} = 12.870 +0.344
\log (V_{Dop}/24.7) - 1.012 \log ({\rm snr}/7.)$ for signal to noise
ratio (per pixel) snr.  
We define two
samples, one for HI column density threshold $\log ({\rm N_{HI}})\geq
13.3$ ($0<z<0.423$, 59 Ly$\alpha$ absorbers) and one for $\log ({\rm
  N_{HI}})\geq 13.1$ ($0<z<0.234$, 47 absorbers).  The column density
of the partial Lyman limit system at $z=0.1671$ was fixed at the value
derived by Prochaska et al. (2003) from FUSE observations.  

In addition to H~I lines, we find many metal absorption lines.
Intervening systems are
at $z$=0.167 (C~II, Fe~II, III; N~II, V; O~I, VI; Si~II, III,
IV); 
$z$=0.1829  (O VI);
$z$=0.3608  (Si III, C IV);
$z$=0.3633  (O VI); $z$=0.4951~(C~III, O~VI).
We find Galactic absorption from Al~II, C~I, II, II$^*$, IV;
Fe~II, N~I, Ni~II, O~I, P~II, S~II, III; Si~II, II$^*$, III, IV.
All Galactic absorption is at $-58 < v < 37$ \kms , so there is no
evidence for any high-velocity clouds.

\section{Ly$\alpha$ forest statistics}

We examine the Ly$\alpha$ forest redshift density $dN/dz$, Doppler
parameter distribution, clustering via the two point correlation
function and the void distribution for the two subsamples with $\log
{\rm N_{HI}}\geq 13.3, 13.1$.

The Ly$\alpha$ forest redshift density has been studied
intensively for over 20 years.  We compare our data against 
recent 
VLT UVES observations (Kim et al. 2002)
and the
literature, and find a consistent value for $14.0<\log {\rm
  N_{HI}}<17.0$ at $z\sim 0.2$ (Fig.~1).  However, for $13.1<\log {\rm
  N_{HI}}<14.0$ the redshift density varies by a factor of two compared to
measurements at $z<0.07$ with GHRS toward 15 extragalactic objects
(Penton et al. 2000).  Cosmic variance may
produce the discrepancy, though 
the path length in this work of 600 $h^{-1}$ Mpc
at $0.020<z<0.234$ ($H_0=70$ \kms\ Mpc$^{-1}$, $\Omega=0.3$, $\Lambda=0.7$)
is large for such a variation.

\begin{figure}[ht]
\epsscale{1.0}
\rule{16cm}{0.2mm}\hfill
\vspace{1mm}
\plotfiddle{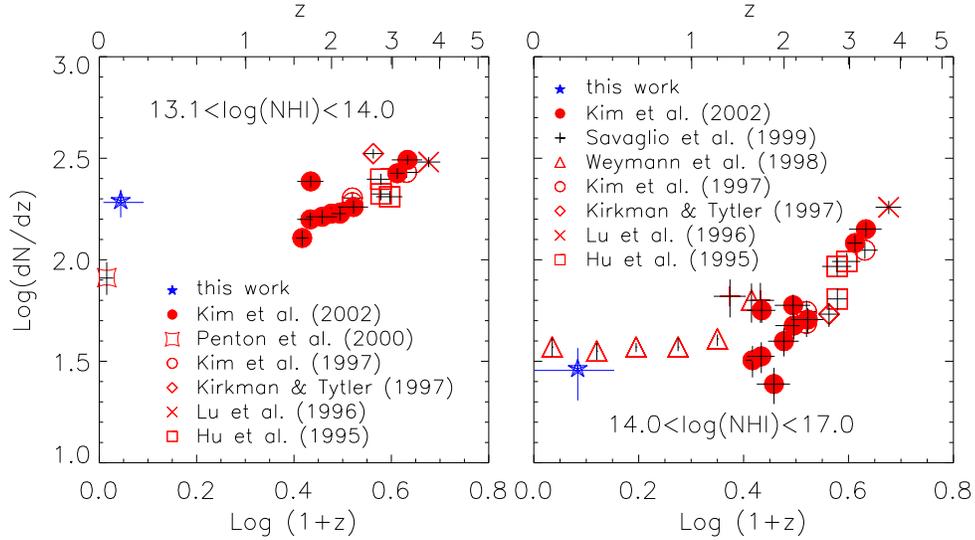}{2.5in}{0.}{60.}{60.}{-210}{-230}
\caption[]{-- Ly$\alpha$ forest redshift density $dN/dz$ for $13.1<\log {\rm N_{HI}}<14.0$ (left)
and $14.0<\log {\rm N_{HI}}<17.0$ (right).  
}
\rule{16cm}{0.2mm}\hfill
\end{figure}

The Doppler parameter distribution for $\log {\rm N_{HI}}\geq 13.3$
(13.1) has mean, median and standard deviation $\langle V_{Dop}
\rangle =$ 48 (45), 47 (45) $\pm 21\, (20)$ \kms .  The large mean arises
from a tail at large values, which likely results from unresolved
blends.


Clustering in the Ly$\alpha$ forest has been studied in a number of
cases (Janknecht et al. 2002 and references therein), with  
weak clustering indicated on velocity scales of $\Delta v < 500$ \kms .  
We use the two point correlation function $\xi(\Delta v)\equiv
[n_{obs}(\Delta v)/n_{exp}(\Delta v)] - 1$ where $n_{obs}$ and
$n_{exp}$ denote the observed and expected numbers of systems in a
relative velocity interval $\Delta v$.  We created 10$^4$ Monte Carlo
simulations weighted in $z$ using $dN/dz \propto (1+z)^\gamma$,
$\gamma = 0.26$ (Weymann et al. 1998).  For each simulation we drew a
number of absorbers from a Poissonian distribution with a mean equal
to the number actually observed in our sample.  We
find a signal for $\log {\rm N_{HI}}\geq 13.2$ at $\Delta v<250$
\kms\ of $\xi(\Delta v) = 1.3$, with 15 pairs observed and $6.49\pm 2.71$
expected (Fig.~2), which has probability $P=0.006$ to be matched or
exceeded by the simulations.  
The
signal weakens at higher $\log {\rm N_{HI}}$, likely suffering from small
number statistics, and at $\log {\rm N_{HI}}\geq 13.1$, perhaps as a
reflection of weaker clustering among weaker perturbations.


\parbox{3.6in}{\epsfxsize=3.6in \epsfbox{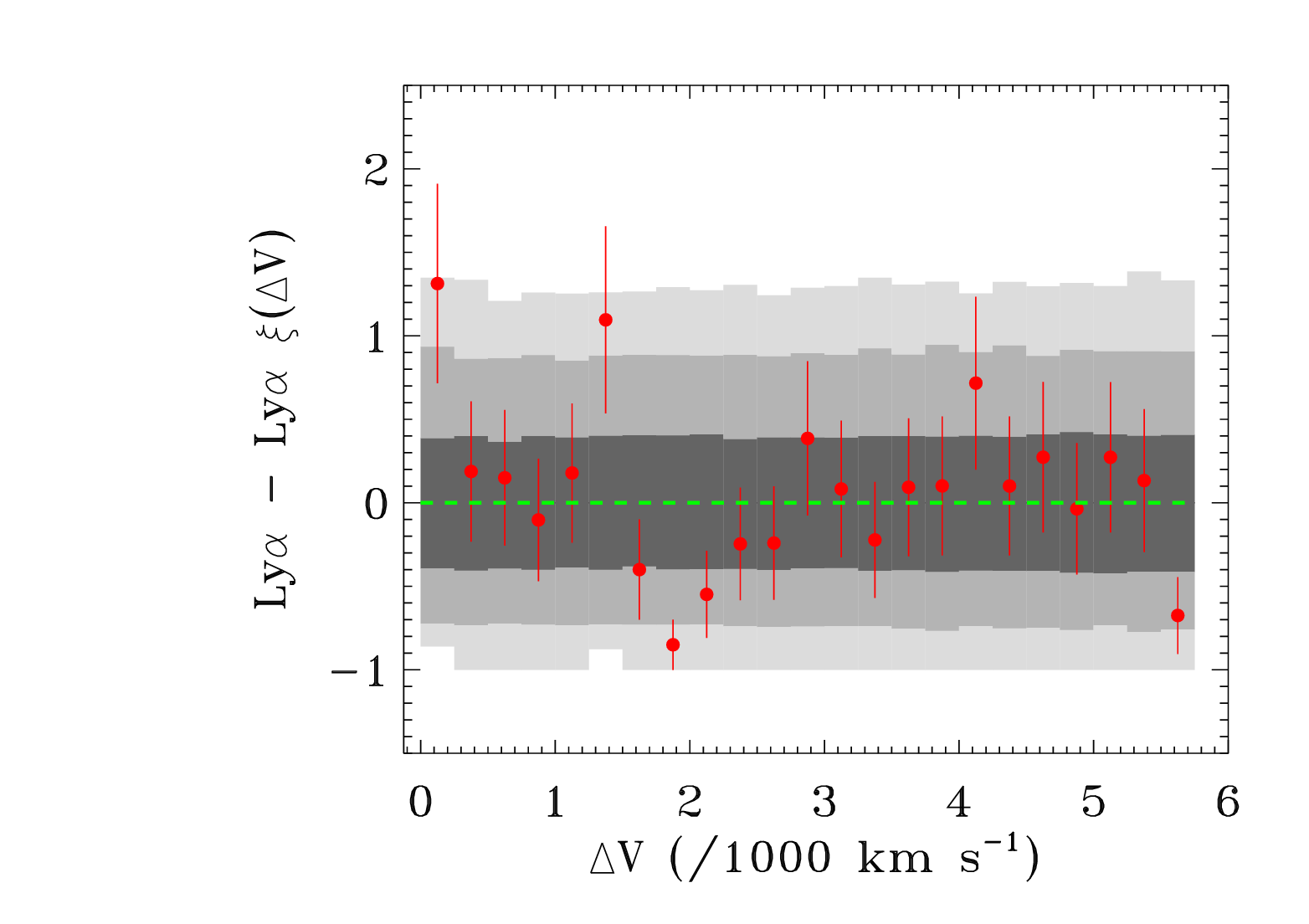}}
\vskip -2in
\hspace{3.8in}
\addtocounter{figure}{1} 
\parbox{2.3in}{\footnotesize Figure \thefigure:-- Two point correlation function for $\log {\rm N_{HI}}\geq 13.2$.
Shaded regions denote 68, 95, 99\% confidence limits from 10$^4$ Monte Carlo simulations.
Error bars show example $1\sigma$ Poissonian errors. }
\vskip 1.3in
\noindent\rule{16cm}{0.2mm}\hfill


We searched for regions in velocity space
devoid of Ly$\alpha$ systems using the same
10$^4$ Monte Carlo simulations, and find evidence for a void at
$0.0320<z<0.0814$ ($\Delta v = 14023$ \kms ) for $\log {\rm
  N_{HI}}\geq 13.3$ with probability $P=0.0005$ to be matched or
exceeded among over $5.7\times 10^5$ $Ly\alpha$ absorber spacings.  For ${\rm
  N_{HI}}\geq 13.2$, we find two voids of similar significance at
$0.0320<z<0.0590$ and $0.1030<z<0.1310$ ($\Delta v = 7744$, 7512 \kms , 
$P=0.005$), while for ${\rm N_{HI}}\geq 13.1$, the lower redshift
void persists ($0.0320<z<0.0590$, $P=0.002$).

\begin{figure}[htb]
\epsscale{1.0}
\rule{16cm}{0.2mm}\hfill
\plotfiddle{pks0405cone.ra.mpc.ps}{1.0in}{-90.}{45.}{45.}{-260}{130}
\plotfiddle{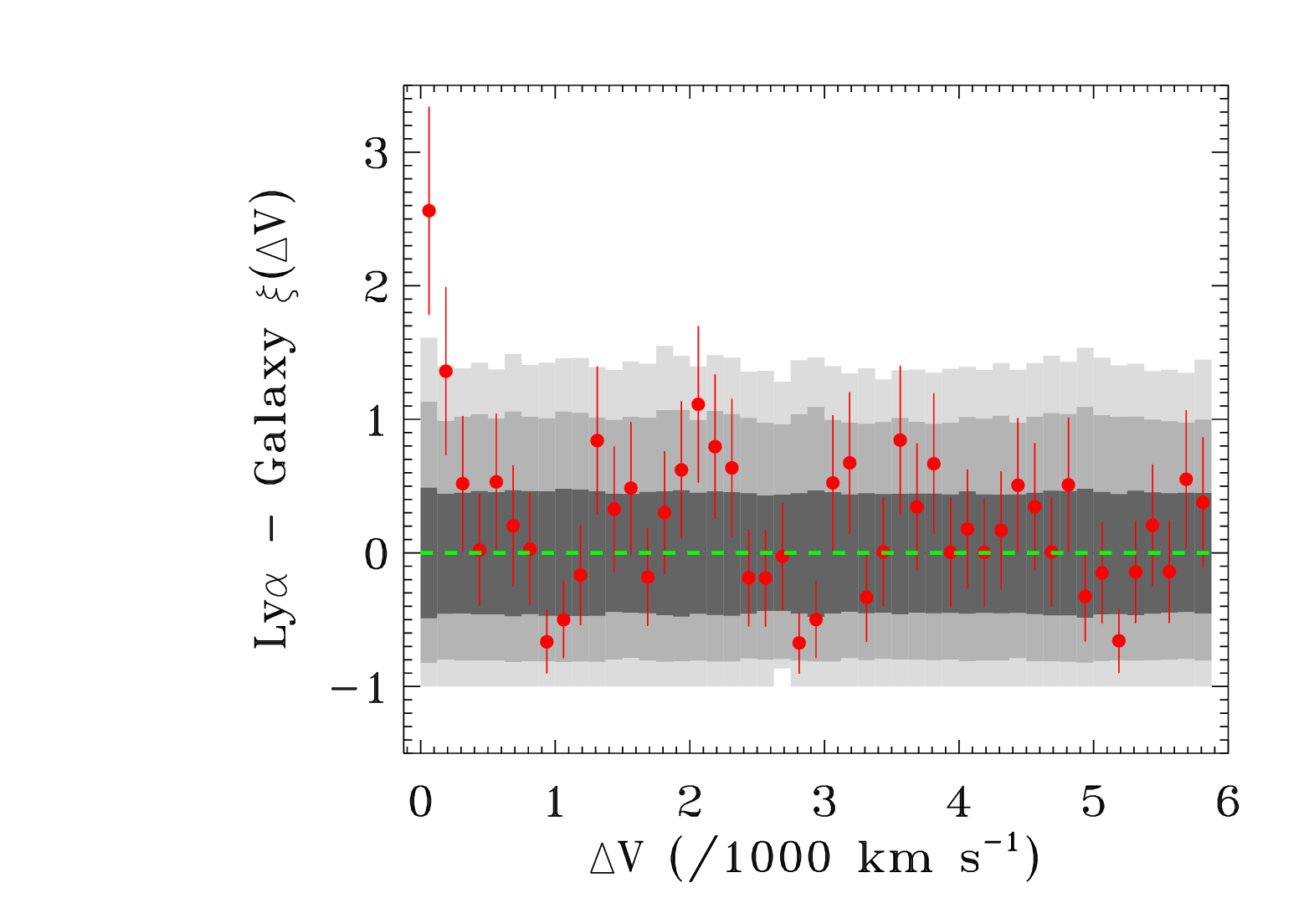}{1.0in}{0.}{46.}{45.}{-60}{-178.3}
\caption[]{-- Ly$\alpha$ absorbers and galaxies toward PKS~0405--123. 
{\bf (a), Left:} A projection of absorbers and galaxies onto the dec plane.
{\it Vertical lines:}
Ly$\alpha$
systems, with length proportional to $\log {\rm N_{HI}}$.
{\it Dashed lines and arrows:} metal absorbers.
{\it Dots:} galaxies.  {\it Thick (thin) line at RA=0:} $z$ limits where
survey is complete to $\log {\rm N_{HI}}=13.1$ (13.3); only absorbers to the
local
completeness limit are shown.  {\it Star:} PKS~0405--123.  {\it Dashed cone:} 
scope corresponding to a 5~arcmin radius.
{\bf (b), Right:} Two point correlation function for Ly$\alpha$ forest and galaxies
within 2 Mpc projected distance, with $\log {\rm N_{HI}}\geq 13.3$.
Shaded regions denote 68, 95, 99\% confidence limits from 10$^4$ Monte Carlo simulations.
Error bars show example $1\sigma$ Poissonian errors.{\hfill\phantom{*}}
}
\rule{16cm}{0.2mm}\hfill
\end{figure}

\section{Ly$\alpha$-galaxy correlations}

Spinrad et al. (1993) and Ellingson \& Yee (1994) 
surveyed for galaxies in a
$10\times 8$ arcmin$^2$ field around PKS~0405--123.  
Data from the NASA Extragalactic Database (NED) and
unpublished data from E. Ellingson provide a list of 45 galaxies at
$z<0.47$ (Fig.~3a) covering $17.75<r<22.94$ ($\langle r \rangle =
20.8\pm 1.1$, $\langle M_r \rangle = -19.9\pm 1.7$ from the distance
modulus for $H_0=70$ \kms\ Mpc$^{-1}$, $\Omega=0.3$, $\Lambda=0.7$),
with redshifts for a further 34 galaxies out to $z=0.78$.  The
spectroscopic completeness is difficult to judge accurately because
some of the objects are stars and there were positional selection
constraints from the multi-slit spectroscopic setup; for $r<22.0$ we
have spectroscopic redshifts for 36 of 213 objects (17\%).


We cross-correlated our galaxy sample with the Ly$\alpha$ forest, 
including all galaxies within 1.5$h^{-1}_{70}$ Mpc in
the local frame.  There is a signal at $\Delta v \leq 250$ \kms\ 
(Fig.~3b, $\xi(\Delta v)=2.0$, 
$5.3\sigma$ significance, 35 pairs observed, $11.8\pm4.3$
expected) which comes from higher $\rm N_{HI}$ absorbers.  The signal
is still detectable for $\log
\rm N_{HI}\geq 14.4$, and persists to $\log
\rm N_{HI}\geq 13.1$, though at $3.0\sigma$ significance
(14 pairs observed, $5.8\pm 2.7$ expected).
There is an apparent cluster around the
$z=0.1671$ partial Lyman limit system, with another around the metal
absorbers at $z\approx 0.36$.  There is also only one galaxy
(at $z=0.0791$) in any of our listed Ly$\alpha$ forest voids.
Bowen et al. (2002) suggested a
correlation between the density of Ly$\alpha$ components along a
sight-line and the volume density of $M_B<-17.5$ galaxies within $\sim
2$ Mpc.  Although our data are highly incomplete, they appear
consistent with this hypothesis.  The field around PKS~0405-123
provides an excellent opportunity to explore the galaxy-Ly$\alpha$
absorber relationship in detail.


\section*{References}
\begin{thebibliography}{99}

\bibitem{Bowen02}Bowen, D. V., Pettini, M., Blades, J. C. 2002, ApJ, 580, 169

\bibitem{Ellingson94}Ellingson, E. \& Yee, H. K. C. 1994, ApJS, 92, 33

\bibitem{Hu95}Hu, E. M., Kim, T.-S., Cowie, L. L., Songaila, A., Rauch, M. 1995, AJ, 110, 1526 

\bibitem{Janknecht02}Janknecht, E., Baade, R., Reimers, D. 2002, A\&A, 391, L11

\bibitem{Kim02}Kim, T.-S. et al. 2002, MNRAS, 335, 555

\bibitem{Kim97}
Kim, T.-S., Hu, E. M., Cowie, L. L.,
Songaila, A., 1997, AJ, 114, 1

\bibitem{Kirkman97} Kirkman, D., Tytler, D., 1997,  AJ, 484, 672

\bibitem{Lu96} Lu, L., Sargent, W. L. W., Womble, D. S.,
Takada-Hidai, M., 1996, ApJ, 472, 509


\bibitem{Penton00} Penton, P. J., Shull, J. M., Stocke, J. T.,
2000, ApJ, 544, 150


\bibitem{Prochaska03}Prochaska, J. et al. 2003, these proceedings

\bibitem{Savaglio99} Savaglio, S., Ferguson, H. C., Brown, T. M.
et al., 1999, ApJ, 515, L5

\bibitem{Spinrad93}Spinrad, H. et al. 1993, AJ, 106, 1

\bibitem{Webb87}Webb, J. K. 1987, PhD thesis, Cambridge University

\bibitem{Weymann98}Weymann, R. J., et al., 1998, ApJ, 506, 1
\end{thebibliography}
\end{document}